\documentclass[apj]{emulateapj}
\slugcomment{{\sc Accepted to ApJ:} August 13, 2012}

\usepackage{graphicx}
\usepackage{amssymb}

\def\aap {{A\&A}}

\def\apjl {{ApJL}}

\def\apjs {{ApJS}}

\newcommand{\HII}{H\,{\sc ii} }
\newcommand{\FLASH}{{\sc flash} }
\newcommand{\ORION}{{\sc orion} }
\newcommand{\Msun}{M_\odot}

\shorttitle{\HII region variability}
\shortauthors{Klassen et al.}

\begin{document}
\bibliographystyle{apj}

\title{\HII region variability and pre-main-sequence evolution}

\author{Mikhail Klassen}
\affil{Department of Physics and Astronomy, McMaster University \\ 1280 Main St.~W, Hamilton, ON L8S 4M1, Canada}
\email{klassm@mcmaster.ca}

\author{Thomas Peters\altaffilmark{1}}
\affil{Institut f\"{u}r Theoretische Physik, Universit\"{a}t Z\"{u}rich \\ Winterthurerstrasse 190, CH-8057 Z\"{u}rich, Switzerland}

\and

\author{Ralph E.~Pudritz\altaffilmark{2}}
\affil{Department of Physics and Astronomy, McMaster University \\ 1280 Main St.~W, Hamilton, ON L8S 4M1, Canada}

\altaffiltext{1}{Zentrum f\"{u}r Astronomie der Universit\"{a}t Heidelberg, Institut f\"{u}r Theoretische Astrophysik, Albert-Ueberle-Str. 2, D-69120 Heidelberg, Germany}
\altaffiltext{2}{Origins Institute, McMaster University, 1280 Main St.~W, Hamilton, ON L8S 4M1, Canada}

\begin{abstract}
Recent observations and simulations have suggested that \HII regions around massive stars may vary in their size and emitted flux on timescales short enough to be observed. This variability can have a number of causes, ranging from environmental causes to variability of the ionizing source itself. We explore the latter possibility by considering the pre-main-sequence evolution of massive protostars and conducting numerical simulations with ionizing radiation feedback using the \FLASH AMR hydrodynamics code. We investigate three different models: a simple ZAMS model, a self-consistent one-zone model by Offner et al. (2009), and a model fit to the tracks computed by Hosokawa \& Omukai (2009). The protostellar models show that hypercompact \HII regions around massive, isolated protostars collapse or shrink from diameters of 80 or 300 AU, depending on the model choice, down to near absence during the swelling of stellar radius that accompanies the protostar's transition from a convective to a radiative internal structure. This occurs on timescales as short as $\sim$3000 years.
\end{abstract}


\section{Introduction}

Massive stars dominate their birth environments through a set of powerful feedback processes, the full implications of which are not completely understood. Among their various feedback mechanisms is the formation of \HII regions---bubbles of hot ($10^4$ K), ionized gas that expand into the colder ($10^2$ K) surrounding medium. These regions are fueled by the prodigious amounts of UV radiation emitted by massive stars \citep{Hoare+2007,Beuther+2007}. By heating and ionizing the gas around them, massive stars alter their birth environments and provide detectable observational signatures.

\HII regions may contribute to the destruction of molecular clouds \citep{Keto2002,Keto2003,Keto2007,Matzner2002}, helping to set cloud lifetimes. They are observable by their radio continuum emission \citep{MezgerHenderson1967}, or by their recombination lines (e.g.\ \citet{WoodChurchwell1989} use the H76$\alpha$ line). Even Young Stellar Objects (YSOs) emit enough UV radiation to begin ionizing the gas around them \citep{Bik+2005,Churchwell2002}, which might teach us something about the lives of massive protostars before they reach the main sequence.

More recently, observations have shown time variability in \HII regions \citep{Franco-HernandezRodriguez2004,Rodriguez+2007,Galvan-Madrid+2008,gomezetal08}. \citet{Franco-HernandezRodriguez2004} have suggested that such observed time-variability may be due to the changes occurring in the source of the ionizing radiation, though it may also be due to increased absorption in the rapidly-evolving core of the nebula.

Three-dimensional collapse simulations of massive star formation that included feedback by ionizing radiation showed time variability with changes in size and flux very similar to the observations \citep{Peters2010a,Klassen+2012}. The origin of this time variability is the strong accretion flow around the massive star, which is dense enough to shield the ionizing radiation. As a consequence, the \HII region fluctuates between extended and trapped states as long as the star is embedded in a sufficiently strong accretion flow. In the simulations, the flickering stops when companion stars form in the gravitationally unstable accretion flow around the central massive star and intercept material that would otherwise be accreted onto the high-mass star, a process we call ``fragmentation-induced starvation'' \citep{Peters2010c}. Then the accretion flow becomes weak enough that the \HII region can isolate the forming high-mass star and terminate the accretion process entirely. Magnetic fields, which provide additional support against gravitational collapse and reduce the number of companion stars formed, do not seem to directly affect the \HII region variability \citep{Peters2011}.

The time variability during the accretion phase can lead to morphological changes of the \HII region appearance over timescales as short as $\sim 10$ years and appears able to resolve the lifetime problem for ultracompact \HII regions \citep{WoodChurchwell1989,Peters2010b}. With the aid of a statistical analysis of the synthetic radio continuum observations of the \HII regions formed in the simulations, \citet{Galvan-Madrid+2011} predicted that 7\% of ultracompact and hypercompact \HII regions should have detectable flux increments, and 3\% should have detectable flux decrements when observed at two epochs separated by about 10 years. Particularly interesting are shrinking \HII regions, for which \citet{Galvan-Madrid+2011} found that flux decrements by at least 10\% are twice as likely as flux decrements by at least 50\% for this time interval.

In general, however, \HII region variability has two possible causes. The first is the chaotic gas motions around a forming star, where denser, optically thick gas can intercept stellar radiation and causing the shadowed gas to neutralize \citep{Peters2010a}. The second possible source of \HII region variability is due to the star itself, as already noted by \citet{Franco-HernandezRodriguez2004}. Pre-main-sequence stars undergo changes during their early evolution, sometimes swelling and other times contracting \citep{YorkeBodenheimer2005}. This can potentially occur on even short time scales ($<2000$ years). Changes in the surface temperature of a star will affect its UV flux, which in turn affects the size of the \HII region.

The connection between protostellar variability and \HII region variability has not yet been explored. To do so, simulations must be equipped with good protostellar models. We ask: can early variability of an \HII region be traced directly to the pre-main-sequence evolution of the massive protostar itself? Models of prestellar evolution have been investigated by \citet{PallaStahler1991}, \citet{PallaStahler1992}, \citet{Nakano2000}, \citet{McKeeTan2003}, \citet{Offner2009} and \citet{HosokawaOmukai2009}, among others, but never investigating the \HII region variability that could result from a massive evolving protostar.

In a previous paper \citep{Klassen+2012}, we simulated the effects of using prestellar models in simulations of clustered massive star formation with initial conditions as in \citet{Peters2010a}. Fluctuations in \HII region size and morphology due to the environmental factors were unaffected by the choice of prestellar model, as might be expected. The prestellar model we implemented, however, did show a delayed onset of major heating and ionization in the cluster.

In the present paper, we investigate the effect of pre-main-sequence evolution on the ionization feedback of a massive protostar with radiation-hydrodynamic simulations that follow the formation and expansion of an \HII region around a single, isolated star embedded in a homogeneous environment. Our reasoning for resorting to a simple model calculation instead of running a full collapse simulation as in \citet{Klassen+2012} is twofold. First, it allows us to separate time variability of the \HII region due to absorption of ionizing radiation by the accretion flow from time variability caused by structural changes in the protostar. In a realistic simulation, both effects would occur simultaneously, making it difficult to assess their relative importance. Second, the prestellar model by \citet{Offner2009}, which was used in our collapse simulations \citep{Klassen+2012}, represents the swelling phase of the more detailed calculations by \citet{HosokawaOmukai2009} only very crudely. Following the behavior in \citet{HosokawaOmukai2009} more closely allows for a better modeling of the time variability. However, accretion rates in collapse simulations are strongly fluctuating \citep{Peters2010a,Klassen+2012}, as opposed to the constant accretion rates considered by \citet{HosokawaOmukai2009}.

Thus, in order to create a realistic model for the time variability caused by pre-main-sequence evolution of the high-mass protostar, we are forced to use idealized simulations with a constant accretion rate as the ones presented in this paper. The work is further idealized because we deliberately only explore constant accretion rates, and thus neglect any time dependence in the HII region due to changing accretion. As a result of this simplicity, we are able to present a clear comparison of three sub-grid models for protostellar effects.

It is at present not possible to run a detailed pre-main-sequence evolution code along with a hydrodynamical simulation, so that our approach of investigating the two aspects of the problem separately appears to be the only solution.

We also stress that we cannot, in general, obtain the time dependence of the \HII region radius by analytical arguments. The key point is that the radius of the protostar may grow or contract depending on the stellar evolutionary stage, which in turn also depends on the accretion history. Since the ionizing flux grows initially during the \HII region expansion, both an enhanced ionizing flux and the thermal pressure of the already ionized gas work together to extend the \HII region radius, so that the \HII region can neither be modeled by a Str\"{o}mgren sphere nor by a standard D-type ionization front. Similar considerations hold for the shrinking phase, in which the ionized gas partially recombines.

Generally, a changing stellar radius affects the surface temperature, which in turn controls the UV flux. This means that if the effective temperature can change significantly over a short period of time, changes in the observed flux or size of \HII regions may signal evolutionary changes taking place in the stars themselves, whose ionizing radiation is driving the region.

Our numerical approach is described in Section \ref{sec:numerical_methods}. In Section \ref{sec:results} we show our results for the evolution of a single massive protostar in a uniform medium, accreting matter at a constant rate, comparing the effects of different accretion rates and different prestellar models. Our assessment of these effects is discussed in Section \ref{sec:discussion}, with some considerations of the observability of \HII region variability. Our findings are summarized in \ref{sec:conclusion}.

\section{Numerical methods}\label{sec:numerical_methods}

We perform numerical simulations using the \FLASH hydrodynamics code \citep{Fryxell2000} in its version 2.5, which solves the hydrodynamics equations on an adaptive, Eulerian grid. Radiative transfer is handled using a raytracing scheme developed by \citet{Rijkhorst} and extended by \citet{Peters2010a}. We fix a single ``star'' at the centre of the simulation box to serve as the source of radiation and evolve a set of stellar parameters based on the models described in Section \ref{sec:protostellar_models} while holding the accretion rate fixed.

The protostellar model provides the radius and the luminosity, from which we can obtain the surface temperature of the star, and hence the fraction of photons with energies capable of ionizing the surrounding the gas. This is not an analytical calculation, but a time-dependent simulation that allows us to compute the flux of ionizing photons, which the radiative feedback code then uses to heat and ionize the gas within the simulation. This allows for a self-consistent calculation of the size of the \HII region and its co-evolution with the protostar.

We perform a grid of simulations with different accretion rates and protostellar models. The three protostellar models that we selected were (1) a purely zero-age main sequence (ZAMS) estimate based on a precomputed table of ZAMS values, (2) a one-zone model based on \citet{Offner2009} and also used in \citet{Klassen+2012}, and (3) a fit to the results from detailed stellar structure calculations performed by \citet{HosokawaOmukai2009}. In all but the ZAMS case, we tested accretion rates of $\dot{M} = 10^{-3}$, $10^{-4}$, and $10^{-5}$ $\Msun$/yr. For the ZAMS case, we only tested $\dot{M} = 10^{-3} \Msun$/yr. For the other two models we have included simulations of $\dot{M} = 10^{-5} \Msun$/yr  for comparison, even though rates higher than $10^{-4} \Msun$/yr are necessary to form a massive star \citep{HosokawaOmukai2009}. Table \ref{table:summary_of_runs} summarizes our simulations.


\begin{deluxetable}{cccc}
\setlength{\tabcolsep}{2pt}
\tabletypesize{\scriptsize}
\tablecaption{Runtime parameters of the single accreting star simulations}
\tablenum{1}
\tablehead{\colhead{Run} & \colhead{Protostellar Model} & \colhead{Accretion Rate} & \colhead{Grid Resolution} \\
\colhead{} & \colhead{} & \colhead{} & \colhead{} }
\startdata
1a &    \citet{Offner2009}  &  $10^{-3} \Msun$/yr  &  20 AU \\
1ah &   \citet{Offner2009}  &  $10^{-3} \Msun$/yr  &  10 AU \\
1b &    \citet{Offner2009}  &  $10^{-4} \Msun$/yr  &  20 AU \\
1c &    \citet{Offner2009}  &  $10^{-5} \Msun$/yr  &  20 AU \\
2i &    Interpolation to ZAMS Table      &  $10^{-3} \Msun$/yr  &  20 AU \\
3a &    Interpolation to H\&O Radius &  $10^{-3} \Msun$/yr  &  20 AU \\
3ah &   Interpolation to H\&O Radius &  $10^{-3} \Msun$/yr  &  6.5 AU \\
3b &    Interpolation to H\&O Radius &  $10^{-4} \Msun$/yr  &  20 AU \\
3c &    Interpolation to H\&O Radius &  $10^{-5} \Msun$/yr  &  20 AU \\
\enddata
\label{table:summary_of_runs}
\end{deluxetable}

\subsection{Protostellar models}\label{sec:protostellar_models}

In \citet{Klassen+2012} we described the implementation of a protostellar evolution module for \FLASH based on the \citet{Offner2009} model implemented in the \ORION code, which we simply called the ``Evolving Protostar'' model. The module considers protostars of mass exceeding 0.1 $\Msun$ and evolves their stellar parameters self-consistently with the simulation. The most important of these parameters is the stellar radius. During the pre-main-sequence evolution of a massive protostar, there comes a point when the stellar structure transitions from a convective core to a radiative core. This is accompanied by a swelling of the stellar radius. The \citet{Offner2009} model estimates a swelling factor of 2.

We compare our results to a ZAMS model for the protostellar radius used by \citet{Peters2010a}. In this model, it is assumed that the radius of the protostar is equal to the radius of an equivalent-mass star on the main sequence. We found that such an approximation for protostars will underestimate the radius by about an order of magnitude \citep{Klassen+2012}. Consequently, estimations of the surface temperature and ionizing luminosity will be strongly overestimated. There is also no variability in the radius besides a steady increase in size as the star accretes matter. No structural transitions in the star are accounted for and the radius never contracts. To implement this model in our simulations, we use tabulated values of the mass, luminosity and effective temperature for a main suquence star. We interpolate within the table based on the current mass of the star and compute the radius from the luminosity and temperature.

The third model we employ is based on the results from \citet{HosokawaOmukai2009} (henceforth H\&O). In their paper, they investigate the evolution of a protostar under high accretion rates through detailed stellar structure calculations. By contrast, rather than computing stellar structure \citet{Offner2009} consider the total energy evolution of a polytrope. One can calibrate a one-zone model to the \citet{HosokawaOmukai2009} calculations, as they describe in Appendix C. However, the accretion rates that we test correspond to the published tracks in H\&O, so we allow our protostellar radius to follow the appropriate track using a fitting function.

\subsection{Initial conditions}\label{sec:initial_conditions}

\begin{figure*}
\centering
\includegraphics[width=160mm]{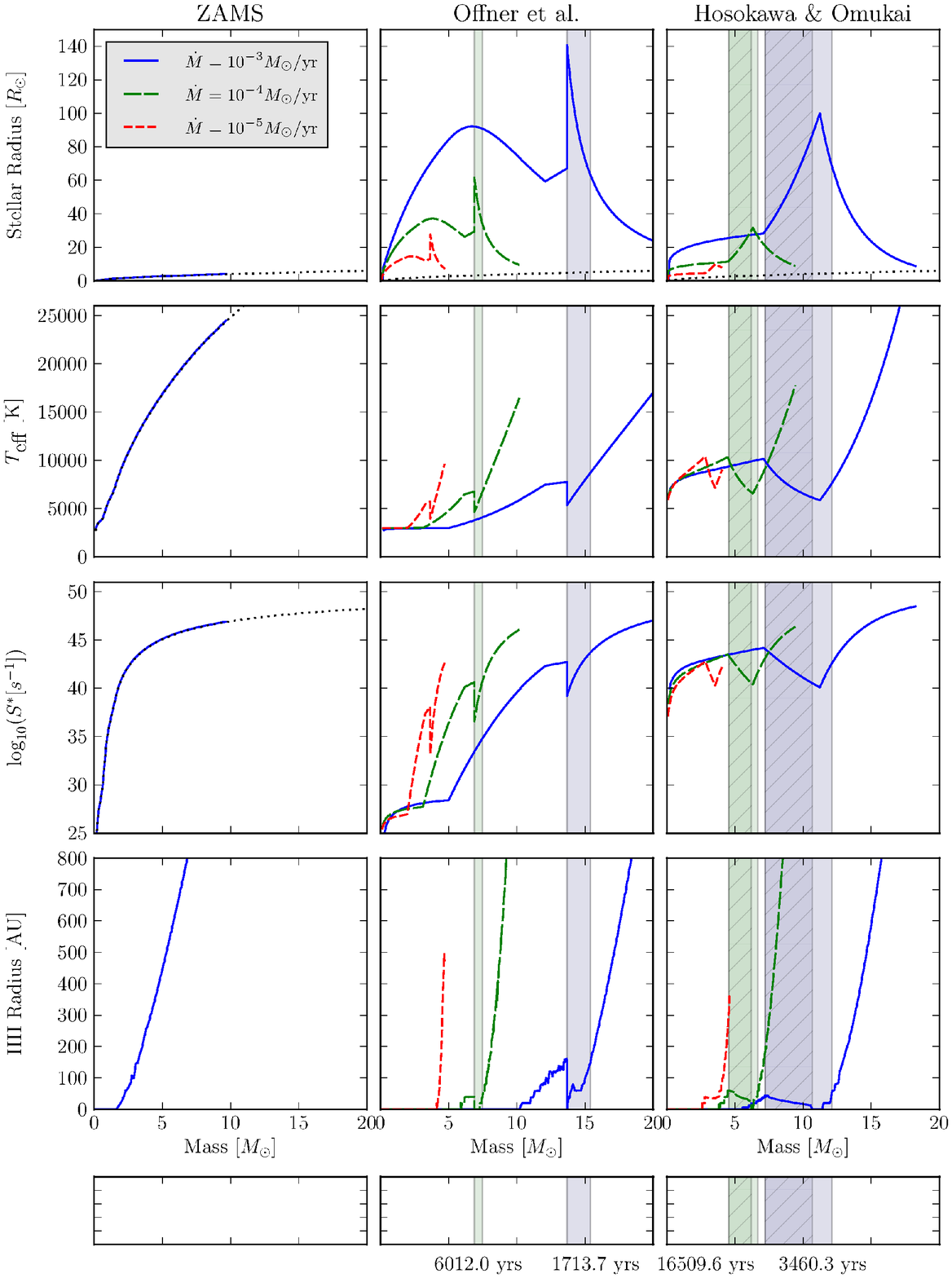}
\caption{Results of a suite of simulations involving a single, isolated protostar accreting mass at either $10^{-3}$, $10^{-4}$, or $10^{-5} \Msun$/yr, and evolving in stellar radius (top row) according to one of three models: an equal-mass ZAMS equivalent (first column), a one-zone model based on \citet{Offner2009} (centre column), or following the results from the stellar structure calculations by \citet{HosokawaOmukai2009} (right column). For each of these models, we also show the surface temperature $T_{\textrm{eff}}$, the ionizing luminosity $S^*$, and the radius of the \HII region surrounding the protostar. Dotted lines show the ZAMS values. The shaded vertical bars indicate the span in the evolution of the protostar between the point where the \HII region has reached its maximal pre-collapse size, and the point post-collapse when the previous size has been reinstated. In the Hosokawa \& Omukai case, the hatched region indicates the span when the \HII region is collapsing. The duration of this collapse is indicated at the bottom of the figure in years. In the Offner et al.~case, the collapse is instantaneous, so the durations at the figure bottom indicate the timescale for regrowth.}
\label{fig:table_star_props}
\end{figure*}

The advantage of the radiative feedback code we employ lies in its ability to produce realistic \HII regions. In a cluster environment, it is difficult to measure the effect of the choice of model on \HII region size because the environment is highly tumultuous even without any initial turbulence or magnetic fields \citep{Peters2010a, Klassen+2012}. The \HII regions formed in such an environment are highly amorphous and time-variable. Here, we want to isolate the effects of the protostellar model to gauge its impact on \HII region formation. To explore this interesting possibility, we select a simulation box of side length 0.1 pc filled with uniform, isothermal molecular gas at 10K and density $\rho = 1 \times 10^{-17}$ g cm$^{-3}$. This is not a collapse calculation and self-gravity of the gas is switched off. We do not consider any initial turbulence, nor magnetic fields, nor do we include radiation pressure.

The radiative feedback code, which is described in greater detail in \citet{Peters2010a}, is a hybrid-characteristics raytracing scheme that computes gas heating with grey atmosphere assumptions, and gas ionization, based on the computed flux of ionizing photons from stars modeled as blackbodies.

A single star is fixed at the centre of this simulation volume whose mass is set to increase at the artificially imposed accretion rate $\dot{M}$, which is held fixed throughout the simulation and is treated as a parameter. By artificially fixing the accretion rate in this way, we can directly connect \HII region variability with the evolution of stellar properties, rather than have it be due to fluctuations in accretion rate or environmental changes as was the case in \citet{Klassen+2012}.

We ran simulations with various prestellar models and accretion rates in order to compare their effect on \HII region formation. The various simulations are summarized in Table \ref{table:summary_of_runs}. We repeated a portion of the \citet{Offner2009} and \citet{HosokawaOmukai2009} models with $\dot{M} = 10^{-3} \Msun$/yr, this time at a high spatial resolution of 10 AU and 6.5 AU minimum grid cell size, respectively. This better captured the behaviour during the transition in stellar structure and swelling of the stellar radius. The high-resolution data was then combined with the lower resolution data.

Each simulation was run for long enough to capture all the major transitions. In most cases, we followed the evolution of the star until it had almost reached the main sequence (see Figure \ref{fig:table_star_props}).

\section{Results}\label{sec:results}

The data from these simulation runs are presented in Figure \ref{fig:table_star_props}, which consists of a series of panels, with the three columns showing each of the three prestellar models we tested.

The rows of panels in this figure show the key variables that we measured or calculated and which describe the evolution of the protostar. The first row shows the stellar radius in solar units. We see how ZAMS model shows nothing but a monotonic increase in stellar radius with increasing mass, while the two prestellar models show a radius that sometimes swells and sometimes contracts. This swelling and contracting of the stellar radius is the key to affecting the size of \HII regions and we trace the effects of this evolution down through the rows of this figure. For comparison, in the first row we show the ZAMS radius in each panel using a thick dashed line.

In the second row, we show calculations of the effective surface temperature of the protostar. This is computed using the equation
\begin{equation}\label{eqn:Teff}
T_{\textrm{eff}} = \left( \frac{L_{\textrm{int}}}{4 \pi \sigma R^2} \right)^{1/4} \, ,
\end{equation}
which depends on the intrinsic luminosity of the protostar $L_{\textrm{int}}$ and the stellar radius $R$. The luminosity in the first case is estimated as in \citet{Offner2009} as
\begin{equation}\label{eqn:Lint}
L_{\textrm{int}} = \textrm{max}(L_{\textrm{ms}} ,4 \pi \sigma R^2 T_{\textrm{H}}^4) \, ,
\end{equation}
that is, the greater of either the main sequence luminosity or the luminosity of a Hayashi-track star with temperature $T_{\textrm{H}} = 3000$ K. We use this same estimate of luminosity for our H\&O comparison. The intrinsic luminosity in the ZAMS case is calculated by interpolation to tabulated values.

These panels show that when the radius of a star swells during the transition to a radiative core structure, the surface temperature drops by several thousand Kelvin. A ZAMS model misses this effect entirely. Capturing this effect in the surface temperature is crucial in correctly determining how the ionizing flux $S^*$, shown in row 3, changes during pre-main-sequence evolution.

The ionizing flux of photons from the protostar is calculated by integrating a blackbody for frequencies above the ionization threshold of 13.6 eV to find the number of ionizing photons per second. This number depends on the surface temperature of the protostar. The prestellar models have drops in their surface temperature, which correspond to drops in ionizing flux. In the case of $\dot{M} = 10^{-3} \Msun$/yr, the \citet{Offner2009} model has a flux decrement of just over 3 orders of magnitude (from $5.3\times10^{42}$ to $1.7 \times 10^{39}$ s$^{-1}$), while the H\&O model shows a flux decrement of just over 4 orders of magnitude (from $1.6 \times 10^{44}$ to $1.2 \times 10^{40}$ s$^{-1}$). The other main difference is that in the former case the drop is instantaneous, whereas the H\&O model show this taking place over 4000 years.

The ZAMS model has only a monotonically increasing ionizing flux and likely also overestimates the ionizing flux for much of the pre-main-sequence lifetime of the star.

Finally, we measured the size of the \HII region by evaluating the ionization fraction along a line starting at the centre of the simulation box, where the particle is located, out along one of the axes to the edge of the simulation box. Sampling the ionization fraction along this axis, we can determine the radius at which the gas transitions from fully ionized to neutral. We define the radius of the \HII region as the point where the ionization fraction $X$ is 0.5, which is found using an inerpolation of the underlying sampling points.

\begin{figure*}
\centering
\includegraphics[width=160mm]{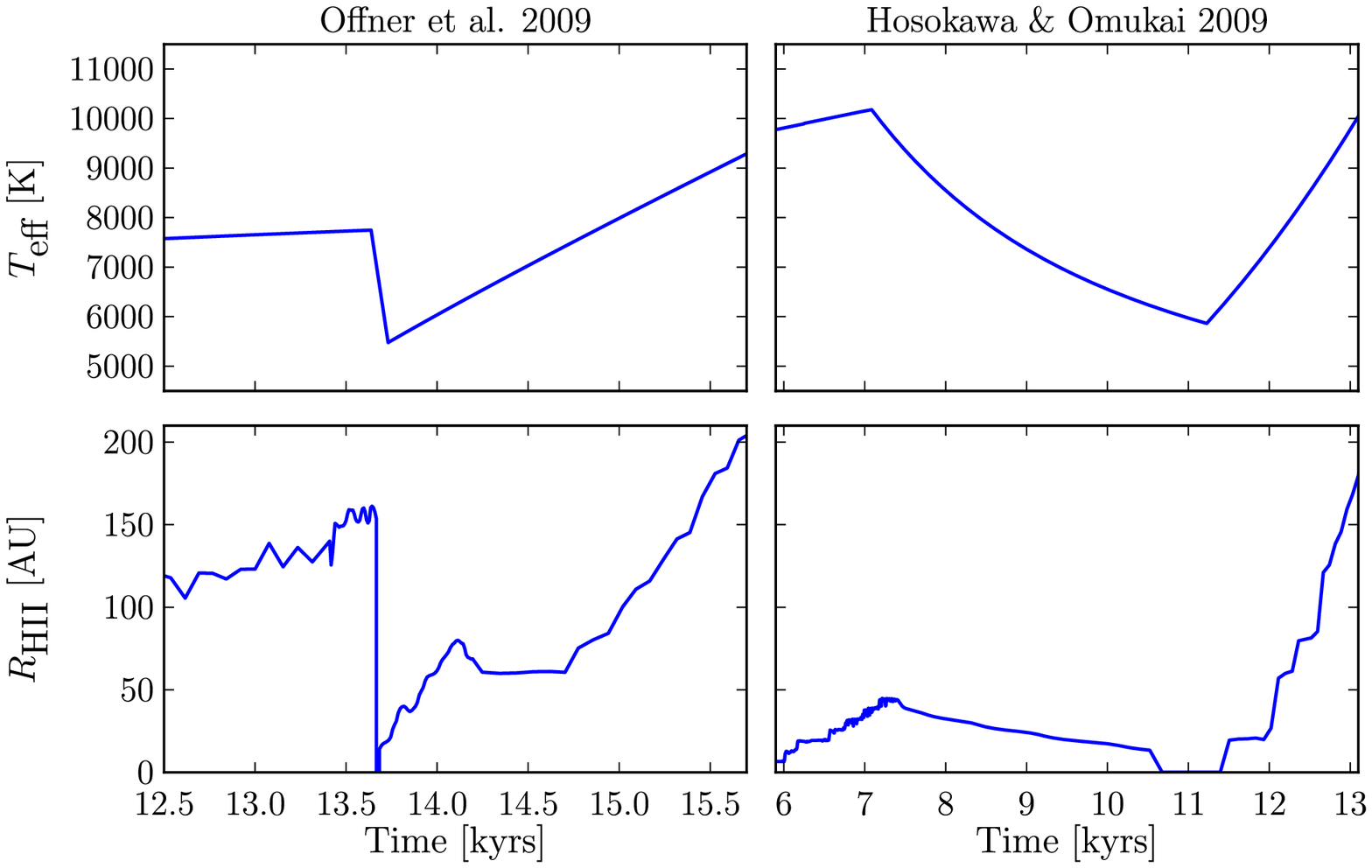}
\caption{A blown-up view of the time of collapse of the \HII region (bottom row) around the protostar. Left panels: the one-zone model by \citet{Offner2009}; right panels: the \citet{HosokawaOmukai2009} data, both at an accretion rate of $10^{-3} \Msun$/yr. The top row shows the concurrent effective surface temperature of the protostar. The small-scale variability of $R_{\textrm{\HII}}$ is partially a numerical effect caused when the \HII region radius is measured in the post-processing of the simulation data and partially a result of small density perturbations. It is on the order of the grid resolution.}
\label{fig:hires_HII}
\end{figure*}

The fourth row of panels in Figure \ref{fig:table_star_props} shows the key result that ties together all the previous findings. Here we show the evolving radius of the \HII region that forms around protostar. The prestellar models of \citet{Offner2009} and H\&O, for accretion rates $\dot{M} \geq 10^{-4} \Msun$/yr, show an \HII region that collapses or shrinks when the ionizing flux is diminished. The loss of ionizing flux is ultimately tied back to the stellar radius, which swells during the pre-main-sequence evolution of the star. For an accretion rate of $\dot{M} = 10^{-5} \Msun$/yr, there was no collapse of the \HII region. In the H\&O case, the growth of the \HII region at this accretion rate appears stagnated for about 100 kyr before it continues to grow.

The temporary loss of the \HII region is more dramatic for the higher accretion rates. The region disappears completely, falling to near zero from a radius of 150 AU in the Offner model, then quickly rebuilding. In the H\&O model, the \HII radius falls to near zero from 40 AU and remains undetectable for about 1700 years.

We added shaded vertical bars that run through Figure \ref{table:summary_of_runs} and mark the key transitions in the prestellar models for accretion rates greater than $10^{-4} \Msun$/yr. The lengths of time that each pair corresponds to are shown at the very bottom of the figure. In the \citet{Offner2009} model, the shaded regions mark the timescale of \HII region collapse and recovery to its pre-collapse size. This destruction and regrowth takes about 1700 years for $\dot{M} = 10^{-3} \Msun$/yr, and about 6000 years for $\dot{M} = 10^{-4} \Msun$/yr.

In the H\&O runs, the shaded region begins when the \HII region has reached its maximal pre-collapse size and ends at the point of full recovery. The first part of the shaded region is hatched and ends where the \HII region has shrunk to zero. The time taken for the region to disappear is indicated at the bottom of the figure, and is about 3400 years for $\dot{M} = 10^{-3} \Msun$/yr, and about 16500 years for $\dot{M} = 10^{-4} \Msun$/yr.

In all our model simulations, as the protostar undergoes its final contraction towards the main sequence, the \HII region grows steadily until we terminate the simulation.

In Figure \ref{fig:hires_HII} we show a closer view of the collapse of an \HII region around a massive protostar for the \citet{Offner2009} and H\&O models. This figure includes high-resolution data, with minimum grid sizes of about 10AU and 6.5 AU, respectively. Time resolution was also higher, at dt = 8 years and dt = 7 years, respectively. In the Offner et al. model, the \HII region radius reaches a maximal value of 161 AU, while in the H\&O case, it grows to only about 45 AU. We show the concurrent effective surface temperatures in the top row, which shows the H\&O reducing in temperature gradually instead of instantaneously. When the flux of ionizing photons falls due to decreasing surface temperature, the \HII region shrinks away to zero over a period of about 3400 years. Only after the surface temperature begins to heat up again does the \HII region begin to reform and grow, as also seen in Figure \ref{fig:table_star_props}. The one-zone model based on \citet{Offner2009} has an instantaneous drop in effective surface temperature, and also an instantaneous drop in \HII region radius.

\section{Discussion}\label{sec:discussion}

Of the models we tested, the \citet{HosokawaOmukai2009} model is likely the most realistic, since in this one we fit functions to their detailed stellar structure calculations. The model also contains no discontinuous jumps in stellar radius, as the \citet{Offner2009} model does. The timescale for flickering, then, in the nascent \HII region is probably closer to the \citet{HosokawaOmukai2009} results.

Given the tumulteous environments in which massive stars form, variability in the \HII regions seems reasonable. \citet{Galvan-Madrid+2008} reported a substantial ($\sim$45\%) decrease in the radio-continuum flux (6 cm) of an unresolved \HII region around a young massive star over a 5 year period. An analysis of radiation-hydrodynamic simulation data from \citet{Peters2010a} by \citet{Galvan-Madrid+2011} using synthetic radio-continuum observations showed that \HII regions can be highly variable over timescales of 10 to $10^4$ years. \HII regions initially larger than 1000 AU around a massive star lost up to 90\% of their size in a short span of time ($\sim$20 years) during a large accretion event.

Those results indicated that the evolution of ultracompact and hypercompact \HII regions may be more complex that the textbook picture of a spherical ionized regions expanding into a quiescent medium. \HII variability is most likely not due to protostellar evolution, which was suggested as a possibility in \citet{Franco-HernandezRodriguez2004}. The variability we observe in our simulations is of a scale too small to explain those effects.

We confirmed that tumultuous birth environments give rise to massive stars with variable \HII regions in our own simulations \citep{Klassen+2012}, but also experimented with a different protostellar model that better captured the evolution of the protostellar radius under conditions of radidly varying accretion. In this paper, we wanted to revisit the toy model of \HII regions to see to which degree a realistic treatment of pre-main-sequence evolution affects the \HII region formation and evolution.

We decoupled the environmental effects of the molecular gas flow through a turbulent environment during the accretion process, to see whether flickering of the \HII region could also be a purely evolutionary feature of the accreting star. We saw the \HII region around the protostar collapse in a similar way, losing over 90\% of its size in a matter of less than 10 years for the \citet{Offner2009} model, or over about 3.5 kyrs for the \citep{HosokawaOmukai2009} model.

Some earlier discussions, based on highly symmetric, simplified models \citep[e.g.][]{Walmsley1995} suggested that high accretion rates might ``choke off'' \HII region formation inside ``hot cores.'' These hot cores are compact (diameters $\leq 0.1$ pc), dense ($n_{H_2} \geq 10^7$ cm$^{-3}$), and at temperatures $\geq$ 100K \citep{Kurtz+2000}. \citet{Churchwell2002} describes these regions as the precursors to ultracompact \HII regions, but also suggests that rapid accretion may prevent young \HII regions from begin detected. However, it is now clear that massive protostars accrete asymmetrically via disks \citep{ZinneckerYorke2007,Beuther+2007}, with some fraction of this infalling material being launched into outflow cavities created by jets \citep{BanerjeePudritz2007}. The jet intensity will be proportional to the infall rate and will also be accompanied by strong beaming of the radiation field around the star in what has been called the ``flashlight'' effect \citep{YorkeBodenheimer1999,YorkeSonnhalter2002,Krumholz+2005}. Analytic models show that an \HII region can be sustained inside a bipolar outflow cavity despite heavy accretion \citep{TanMcKee2003}. Numerical simulations of non-spherical accretion \citep{Peters2010a,Klassen+2012} demonstrate that \HII regions can expand in spite of strong accretion flows.

Under circumstances such as these, the \HII region would not be choked off, despite the very high accretion rate. To our knowledge, accretion choking out the formation of an \HII region has never been seen in numerical simulations, nor is accretion ever spherically symmetric onto a massive star under realistic conditions. Therefore, it is timely to explore the birth and evolution of hypercompact \HII regions for massive protostars at high accretion rates, even if our spherically symmetric model is highly idealized. We sought to explore the time-evolution of an \HII region around a source whose flux of ionizing photons is varying according the time-evolution of the protostellar radius and surface temperature. No analytical model for the time-variability of \HII regions exists and that time-variability exists whether the morphology is spherically symmetric or not.

Many numerical calculations of protostellar evolution \citep{PallaStahler1991,YorkeBodenheimer2005,HosokawaOmukai2009}, even for first stars \citep{TanMcKee2004}, have shown that the protostellar radius undergoes a swelling phase followed by a contracting phase as the star approaches the main sequence. \citet{Smith+2012} considered the time-evolution of the protostellar radius for first stars and showed that under some conditions the radius can undergo multiple occasions of swelling and contraction on account of a variable accretion rate. We measured the impact of this evolution on the size of the nascent \HII region and showed that fluctuation in the size does indeed occur, but are small enough that they might be very difficult to observe with present radio telescopes.

\subsection{Observational considerations}

Hypercompact \HII regions are nominally defined with sizes of about 30 mpc (6000 AU) or smaller. Understanding their place in the scheme of massive star formation is important \citep{Kurtz2005}. \HII regions, generally, are associated with star clusters, and we discussed their variability in \citet{Klassen+2012}. Higher-resolution surveys will ultimately yield more examples of extremely compact \HII regions associated with individual, massive stars instead of clusters that may tell us more about the dynamics and kinematics of the regions that give rise to a massive star.

\citet{TanMcKee2003} proposed that hypercompact \HII regions would be confined to the outflow cavities created by rapidly-accreting, massive protostars. The faster the acccretion, the higher the outflow rate. In our self-consistent collapsing cluster simulations \citep{Klassen+2012}, we saw accretion rates that at times approached $10^{-3} \Msun$/yr. \citet{HosokawaOmukai2009} also performed their calculations of protostellar evolution with accretion rates up to $10^{-3} \Msun$/yr. 

Our idealized model of a single star accreting at a fixed rate and feeding back into a homogeneous medium isolates the massive protostar from the cluster environment that was studied in \citet{Klassen+2012}. In this isolated environment, we could see the variability of the \HII region, as caused by the protostellar evolution instead of the environment. It was for this reason that the environment was made as simple as possible. Though massive protostars in nature accrete material asymmetrically through a disk, this does not affect the fact that \HII regions will be variable according to their protostellar evolution, which involves a rapid swelling of the stellar radius that drastically lowers the surface temperature and fraction of photos above the ionization threshold of 13.6 eV. This also makes accretion onto the protostar easier.

Under the models we tested, the diameter of \HII regions collapsed at rate of 45 AU/yr (for the Offner et al. 2009 model), and at a rate of 0.03 AU/yr (for the Hosokawa \& Omukai 2009 model), where in both cases the protostar was accreting at a rate of $10^{-3} \Msun$/yr. The Orion Molecular Cloud is located about 460 pc away. In observing the radio continuum at 1.3 cm, the EVLA has an angular resolution of about 8.7 mas, which corresponds to a spatial resolution of about 4 AU at 460 pc, and 17.4 AU at 2 kpc. This suggests that \HII region flickering around massive protostars could be observable, following the \citet{Offner2009} model. The \HII regions would not appear quite as they do in our simulations, given that massive protostars are embedded in dense gas and dusk, but they might be seen via outflows as suggested earlier.

Additionally, masers serve as useful probes of high-mass star-forming environments and can be associated with ultracompact \HII regions \citep{Fish+2005}. VLBI techniques may offer the angular resolution at a few kiloparsecs that ALMA cannot. Multi-epoch observations of the hydroxyl masers around W3(OH) with time baseline of 8 years showed evidence consistent with an expanding ultracompact \HII region \citep{Bloemhof+1992}. Even with archival data, the authors were capable of positioning the sources with sub-milliarcsecond resolution, corresponding to maser positioning to within better than 2 AU at 2.2 kpc. If masers drift is truly associated with \HII region variability and could be detected in the earliest stages of massive star formation, variability on timescales shorter than 10 years could potentially be studied.

\section{Conclusions}\label{sec:conclusion}

Our results show that massive protostars in uniform media emit ionizing radiation capable of forming a hypercompact \HII region. In our simulations using the \FLASH astrophysics code with ionizing feedback and different prestellar models, the pre-main-sequence evolution of protostars can cause early variability in the size of these \HII regions. During pre-main-sequence evolution, the stellar radius swells dramatically when the star transitions to a radiative internal structure. This is accompanied by a cooling of the surface temperature and a drop in the ionizing flux by several orders of magnitude. During this transition, young \HII regions will shrink or collapse until the contracting star has regained its previous ionizing flux. 

Our results show that pre-main-sequence evolution can affect the growth of \HII regions around massive protostars on potentially short time scales. The collapse of the \HII region, initially $\sim$300 AU across in the \citet{Offner2009} model, happened within a single simulation timestep (8 years) when the radius of the star was enlarged during the evolution. The \citet{HosokawaOmukai2009} model saw the radius of the \HII region contract from an initial diameter of $\sim$90 AU to zero over the course of about 3.5 kyr.

The observability of this variability may be possible with the best interferometric radio observations such as with VLBI, the EVLA, or ALMA. Nevertheless, model simulations of hypercompact \HII regions are an important part of understanding the complete picture of massive star formation. Massive protostars are unique in that their UV fluxes are large enough to ionize gas even before the star has reached the main sequence, and it is therefore important that these early states be further explored.

\section*{Acknowledgments}

We thank our anonymous referee for a helpful report. M.K.~acknowledges financial support from the Ontario Graduate Scholarship (OGS) Program. T.P.~acknowledges financial support as a Fellow of the Baden-W\"{u}rttemberg Stiftung funded by their program International Collaboration II (grant P-LS-SPII/18) and through SNF grant 200020\_137896. R.E.P.~is supported by a Discovery Grant from the Natural Sciences and Engineering Research Council (NSERC) of Canada. The \FLASH code was in part developed by the DOE NNSA-ASC OASCR Flash Center at the University of Chicago. This work was made possible by the facilities of the Shared Hierarchical Academic Research Computing Network (SHARCNET: www.sharcnet.ca) and Compute/Calcul Canada.

\bibliography{single_star_accretion_apj}

\label{lastpage}

\end{document}